\pdfoutput=1
\documentclass[final, 10pt, times, twocolumn, twoside]{article}

\usepackage[left=15mm, right=15mm, bottom=15mm, top=15mm]{geometry}
\usepackage[usenames,dvipsnames]{color}
\usepackage[super, sort&compress]{natbib}
\usepackage{mathtools}
\usepackage{amsmath}
\usepackage{amssymb}
\usepackage{enumerate}
\usepackage{comment}
\usepackage{fix-cm}
\usepackage{url}
\usepackage[colorlinks=true, linkcolor=black, citecolor=blue, urlcolor=blue]{hyperref} %for the links
\usepackage[english]{babel}
\usepackage{graphicx}
\usepackage{tabularx}
\usepackage[table]{xcolor}
\usepackage{booktabs} %for the rules
\usepackage{float}
\usepackage[font=footnotesize, labelfont=bf, justification=justified, singlelinecheck=false]{caption} %hyperlink navigates to the top of the figure
\usepackage[switch]{lineno}
\usepackage[bottom]{footmisc}
\usepackage{setspace} %for inter-line
\usepackage{fancyhdr} 
\usepackage{soul} %for highlighting

\makeatletter
\renewcommand{\fnum@figure}{Fig.~\thefigure}
\makeatother
\newcolumntype{L}[1]{>{\raggedright\let\newline\\\arraybackslash\hspace{0pt}}m{#1}}
\newcolumntype{C}[1]{>{\centering\let\newline\\\arraybackslash\hspace{0pt}}m{#1}}
\newcolumntype{R}[1]{>{\raggedleft\let\newline\\\arraybackslash\hspace{0pt}}m{#1}}

\begin{document}

\twocolumn[
  \begin{@twocolumnfalse}

   \ \\ \ \\ \ \\ \ \\
   {\Huge Accurate autocorrelation modeling substantially} \ \\ \ \\
   {\Huge improves fMRI reliability} \ \\ \ \\ \ \\ \ \\

   Wiktor Olszowy*$^1$, John Aston$^2$, Catarina Rua$^1$ \& Guy B. Williams$^1$

   \ \\ \ \\ \ \\

   {\setstretch{1.5} %-inter-line
   \begin{minipage}{0.7\textwidth}
   Given the recent controversies in some neuroimaging statistical methods, we compare the most frequently
   used functional Magnetic Resonance Imaging (fMRI) analysis packages: AFNI, FSL and SPM, with regard to
   temporal autocorrelation modeling. This process, sometimes known as pre-whitening, is conducted in virtually
   all task fMRI studies.
   We employ eleven datasets containing 980 scans corresponding to different
   fMRI protocols and subject populations.
   Though autocorrelation modeling in AFNI is not perfect, its performance is much higher than the
   performance of autocorrelation modeling in FSL and SPM.
   The residual autocorrelated noise in FSL and SPM leads to heavily confounded first level
   results,
   particularly for low-frequency experimental designs.
   Our results show superior performance of SPM's alternative
   pre-whitening: FAST, over SPM's default.
   The reliability of task fMRI studies would increase with more accurate autocorrelation modeling.
   Furthermore, reliability could increase if the packages
   provided diagnostic plots. This way the investigator would be aware of pre-whitening problems.
   \end{minipage} }

   \ \\ \ \\ \ \\ \ \\ \ \\ \ \\ \ \\ \ \\ \ \\ \ \\

   $^1$Wolfson Brain Imaging Centre, Department of Clinical Neurosciences, University of Cambridge, Cambridge, CB2 0QQ, United Kingdom.
   $^2$Statistical Laboratory, Department of Pure Mathematics and Mathematical Statistics, University of Cambridge, Cambridge, CB3 0WB, United Kingdom.
   Correspondence and requests for materials should be addressed to W.O. (email: wo222@cam.ac.uk).

  \end{@twocolumnfalse}
]

\clearpage

\noindent
Functional Magnetic Resonance Imaging (fMRI) data is known to be positively autocorrelated in
time\cite{bullmore1996statistical}. It results from neural sources, but also from scanner-induced
low-frequency drifts, respiration and cardiac pulsation, as well as from movement artefacts not
accounted for by motion correction\cite{lund2006non}. If this autocorrelation is not accounted
for, spuriously high fMRI signal at one time point can be prolonged to the subsequent time points,
which increases the likelihood of obtaining false positives in task
studies\cite{purdon1998effect}.~As a result, parts of the brain might erroneously appear active during an
experiment. The degree of temporal autocorrelation is different across the brain\cite{worsley2002general}.
In particular, autocorrelation in gray matter is stronger than in white matter and cerebrospinal fluid,
but it also varies within gray matter.

\begin{table*}[t]
   \vspace{1mm}
   \caption{\textbf{Overview of the employed datasets.}}
   \footnotesize{
   \begin{tabularx}{\textwidth}{@{}XL{1.9cm}L{2.2cm}C{1.90cm}C{1.29cm}C{0.77cm}C{0.8cm}C{1.65cm}C{1.7cm}C{0.93cm}}
      \textbf{Study} & \textbf{Experiment} & \textbf{Place} & \textbf{Design}  & \textbf{No.}      & \textbf{Field} & \textbf{TR}  & \textbf{Voxel}     & \textbf{No.}    & \textbf{Time}   \\
                     &                     &                &                  & \textbf{subjects} & \textbf{[T]}   & \textbf{[s]} & \textbf{size [mm]} & \textbf{voxels} & \textbf{points} \\ \toprule
         FCP         & resting state       & Beijing        & N/A              & 198               & 3              & 2            & 3.1x3.1x3.6        & 64x64x33        & 225 \\
                     & resting state       & Cambridge, US  & N/A              & 198               & 3              & 3            & 3x3x3              & 72x72x47        & 119 \\
         NKI         & resting state       & Orangeburg, US & N/A              & 30                & 3              & 1.4          & 2x2x2              & 112x112x64      & 404 \\
                     & resting state       & Orangeburg, US & N/A              & 30                & 3              & 0.645        & 3x3x3              & 74x74x40        & 900 \\
         CRIC        & resting state       & Cambridge, UK  & N/A              & 73                & 3              & 2            & 3x3x3.8            & 64x64x32        & 300 \\
         neuRosim    & resting state       & (simulated)    & N/A              & 100               & NA             & 2            & 3.1x3.1x3.6        & 64x64x33        & 225 \\ \midrule
         NKI         & checkerboard        & Orangeburg, US & 20s off+20s on   & 30                & 3              & 1.4          & 2x2x2              & 112x112x64      & 98  \\
                     & checkerboard        & Orangeburg, US & 20s off+20s on   & 30                & 3              & 0.645        & 3x3x3              & 74x74x40        & 240 \\
         BMMR        & checkerboard        & Magdeburg      & 12s off+12s on   & 21                & 7              & 3            & 1x1x1              & 182x140x45      & 80  \\
         CRIC        & checkerboard        & Cambridge, UK  & 16s off+16s on   & 70                & 3              & 2            & 3x3x3.8            & 64x64x32        & 160 \\ \midrule
         CamCAN      & sensorimotor        & Cambridge, UK  & event-related    & 200               & 3              & 1.97         & 3x3x4.44           & 64x64x32        & 261 \\ \bottomrule
   \end{tabularx} }
   \vspace{2mm}
   \ \\
   \noindent
   FCP = Functional Connectomes Project. NKI = Nathan Kline Institute.
   BMMR = Biomedical Magnetic Resonance. CRIC = Cambridge Research into Impaired Consciousness.
   CamCAN = Cambridge Centre for Ageing and Neuroscience. For the Enhanced NKI data, only scans from release
   3 were used. Out of the 46 subjects in release 3, scans of 30 subjects were taken. For the rest,
   at least one scan was missing. For the BMMR data, there were 7 subjects at 3 sessions, resulting in
   21 scans. For the CamCAN data, 200 subjects were considered only.
   \label{studies_table}
\end{table*}

AFNI\cite{cox1996afni}, FSL\cite{jenkinson2012fsl} and SPM\cite{penny2011statistical},
the most popular packages used in fMRI research, first remove the signal at very low frequencies
(for example using a high-pass filter), after which they estimate the residual temporal
autocorrelation and remove it in a process called pre-whitening.
In AFNI temporal autocorrelation is modeled voxel-wise. For each voxel, an
autoregressive-moving-average ARMA(1,1) model is estimated. The two ARMA(1,1) parameters are estimated only
on a discrete grid and are not spatially
smoothed. For FSL, a Tukey taper is used to smooth the spectral density estimates voxel-wise. These smoothed
estimates
are then additionally smoothed within tissue type. Woolrich et al.\cite{woolrich2001temporal} has shown
the applicability of the FSL's method in two
fMRI protocols: with repetition time (TR) of 1.5s and of 3s, and with voxel size
4x4x7 mm$^3$.
By default, SPM estimates temporal autocorrelation globally as an autoregressive AR(1) plus white noise
process\cite{friston2002classical}.~SPM has an alternative approach: \texttt{FAST}, but we know of only three
studies which have used it\cite{todd2016evaluation, BOLLMANN2018152, corbin2018accurate}.
\texttt{FAST} uses a dictionary of covariance components based on exponential covariance functions\cite{corbin2018accurate}.
More specifically, the dictionary is of length $3 p$ and is composed of $p$ different exponential time constants
along their first and second derivatives. By default, \texttt{FAST} employs 18 components.
Like SPM's default pre-whitening method, \texttt{FAST} is based on a global noise model.

Lenoski et al.\cite{lenoski2008performance} compared several fMRI autocorrelation modeling approaches
for one fMRI protocol (TR=3s, voxel size 3.75x3.75x4 mm$^3$).
The authors found that the use of the global AR(1), of the spatially smoothed AR(1) and of the spatially
smoothed FSL-like noise models resulted in worse whitening performance than the use of the non-spatially
smoothed noise models.
Eklund et al.\cite{eklund2012does} showed that in SPM the shorter the TR, the more likely it is to get false positive
results in first level (also known as single subject) analyses. It was argued that SPM often does not
remove a substantial part of the autocorrelated noise.
The relationship between shorter TR and increased false positive rates was also shown for the case when
autocorrelation is not accounted\cite{purdon1998effect}.

In this study we investigate the whitening performance of AFNI, FSL and SPM for a wide variety of fMRI
protocols. We analyze both the default SPM's method and the alternative one: \texttt{FAST}. Furthermore, we
analyze the resulting specificity-sensitivity trade-offs in first level fMRI results, and investigate the
impact of pre-whitening on second level analyses. We observe better whitening
performance for AFNI and SPM tested with option \texttt{FAST} than for FSL and SPM. Imperfect pre-whitening
heavily confounds first level analyses.

\begin{figure*}[!t]
   \begin{center}
   	\includegraphics[width=0.85\textwidth]{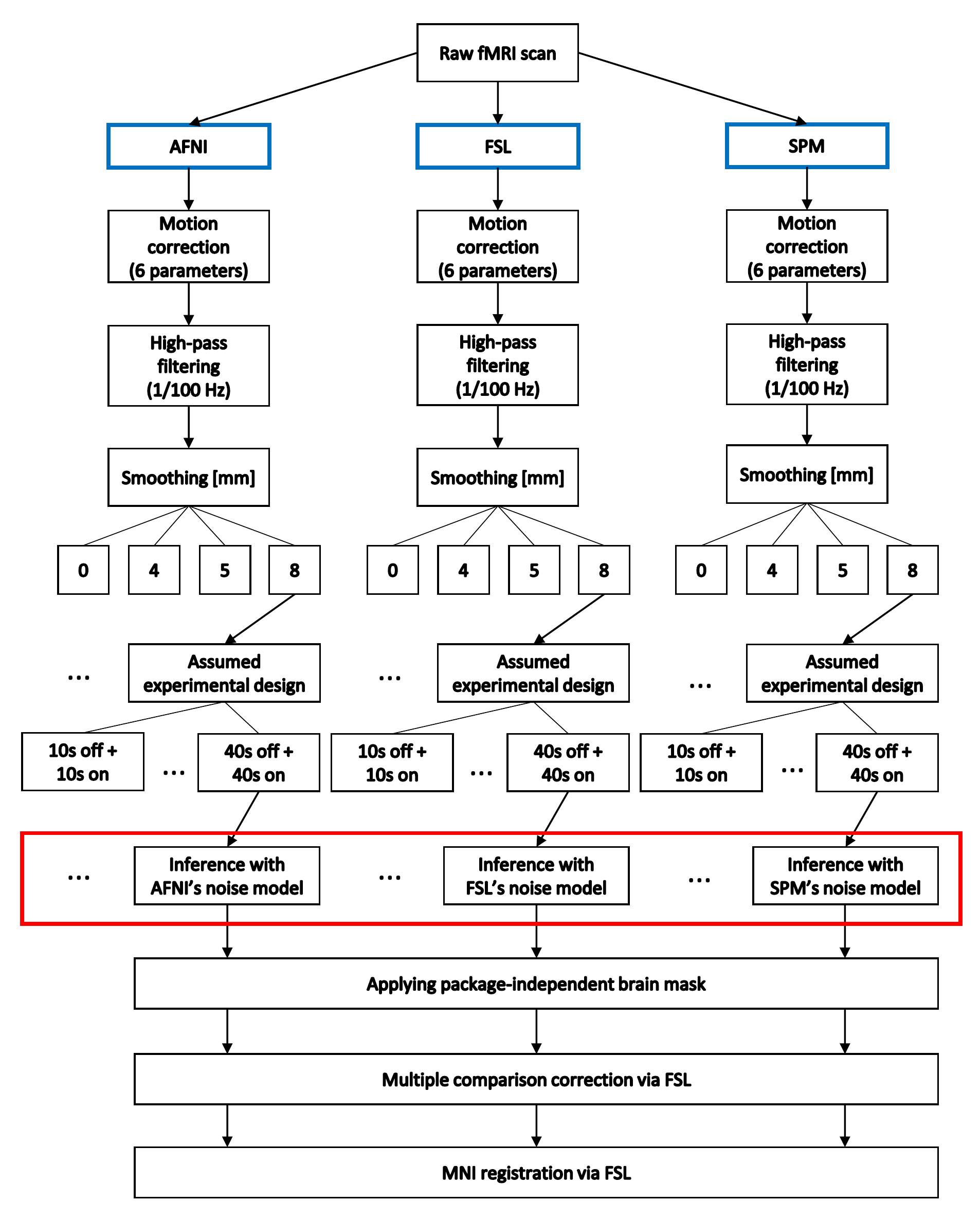}
   \end{center}
   \vspace{-5mm}
	\caption{The employed analyses pipelines. For SPM, we investigated both the default noise model and the
      alternative noise model: \texttt{FAST}. The noise models used by AFNI, FSL and SPM were the only relevant
      difference (marked in a red box).}
   \label{proc_graph}
\end{figure*}

\ \\
\textbf{Methods} \ \\
\textbf{Data.}
In order to explore a range of parameters that may affect autocorrelation, we investigated 11 fMRI datasets
(Table~\ref{studies_table}). These included resting state and task studies, healthy subjects and a patient
population, different TRs, magnetic field strengths and voxel sizes.
We also used anatomical MRI scans, as they were needed
for the registration of brains to the MNI (Montreal Neurological Institute) atlas space.
FCP\cite{biswal2010toward}, NKI\cite{nooner2012nki} and CamCAN data\cite{shafto2014cambridge}
are publicly shared anonymized data. Data
collection at the respective sites was subject to their local institutional review boards (IRBs), who
approved the experiments and the dissemination of the
anonymized data. For the 1,000 Functional Connectomes Project (FCP), collection of the Beijing data was
approved by
the IRB of State Key Laboratory for Cognitive Neuroscience and Learning, Beijing Normal University; collection
of the Cambridge data was approved by the Massachusetts General Hospital partners’ IRB.
For the Enhanced NKI Rockland Sample, collection and dissemination of the data was
approved by the NYU School of Medicine IRB.
For the analysis of an event-related design dataset, we used
the CamCAN dataset (Cambridge Centre for Ageing and Neuroscience,~\url{www.cam-can.org}).
Ethical approval for the study was obtained from
the Cambridgeshire 2 (now East of England - Cambridge Central) Research Ethics Committee.
The study from Magdeburg, ``BMMR checkerboard''\cite{hamid2015quantitative},
was approved by the IRB of the Otto von Guericke University. The study of Cambridge Research into
Impaired Consciousness (CRIC) was approved by the Cambridge Local Research Ethics Committee (99/391).
In all studies all subjects or their consultees gave informed written consent after the experimental
procedures were explained.
One rest dataset consisted of simulated data generated with the \texttt{neuRosim} package in
\texttt{R}\cite{welvaert2011neurosim}. Simulation details can be found in Supplementary Information.

\ \\
\noindent
\textbf{Analysis pipeline.}
For AFNI, FSL and SPM analyses, the preprocessing, brain masks, brain registrations to the 2 mm isotropic MNI
atlas space, and multiple comparison corrections were kept consistent (Fig.~\ref{proc_graph}).
This way we limited the influence of possible confounders on the results.
In order to investigate whether our results are an artefact of the comparison approach used for assessment, we
compared AFNI, FSL and SPM by investigating
(1) the power spectra of the GLM residuals,
(2) the spatial distribution of significant clusters,
(3) the average percentage of significant voxels within the brain mask, and
(4) the positive rate: proportion of subjects with at least one significant cluster.
The power spectrum represents the variance of a signal that is attributable to an
oscillation of a given frequency.
When calculating the power spectra of the GLM residuals, we considered voxels in native space using the same
brain mask for AFNI, FSL and SPM. For each voxel, we normalized the time series to have variance 1 and
calculated the power spectra as the square of the discrete Fourier transform.
Without variance normalization, different signal scaling across voxels and subjects would make it difficult
to interpret power spectra averaged across voxels and subjects.

Apart from assuming dummy designs for resting state data as in recent
studies\cite{eklund2012does, eklund2015empirically, eklund2016cluster}, we also assumed wrong (dummy)
designs for task data, and we used resting state scans simulated using the \texttt{neuRosim} package in
\texttt{R}\cite{welvaert2011neurosim}. We treated such data as null data. For null data, the positive rate
is the familywise error rate, which was investigated in a number of recent studies\cite{eklund2012does, eklund2015empirically,
eklund2016cluster}. We use the term ``significant voxel'' to denote a voxel that is covered by one of the
clusters returned by the multiple comparison correction.

All the processing scripts needed to fully replicate our study are at
\url{https://github.com/wiktorolszowy/fMRI_temporal_autocorrelation}.
We used AFNI 16.2.02, FSL 5.0.10 and SPM 12 (v7219).

\ \\
\noindent
\textbf{Preprocessing.}
Slice timing correction was not performed as part of our main analysis pipeline, since for some datasets
the slice timing information was not
available. In each of the three packages we performed motion correction, which
resulted in 6 parameters that we considered as confounders in the consecutive statistical analysis.
As the 7T scans from the ``BMMR checkerboard'' dataset were prospectively motion corrected\cite{thesen2000prospective},
we did not perform motion correction on them. The ``BMMR checkerboard'' scans were also prospectively distortion
corrected\cite{in2012highly}.
For all the datasets, in each of the three packages we conducted high-pass filtering with frequency cut-off of
1/100 Hz.
We performed registration to MNI space only within FSL. For AFNI and SPM, the results of the multiple
comparison correction were registered to MNI space using transformations generated by FSL.
First, anatomical scans were brain extracted with FSL's brain extraction tool (BET)\cite{smith2002fast}.
Then, FSL's boundary based registration (BBR) was used for registration of the fMRI volumes to the
anatomical scans.
The anatomical scans were aligned to 2 mm isotropic MNI space using affine registration with 12 degrees of
freedom. The two transformations were then combined for each subject and saved for later use in all analyses,
including in those started in AFNI and SPM.
Gaussian spatial smoothing was performed in each of the packages separately.

\ \\
\noindent
\textbf{Statistical analysis.}
For analyses in each package, we used the canonical hemodynamic response function (HRF) model,
also known as the double gamma model. It is implemented the same way in AFNI, FSL and SPM: the response
peak is set at 5 seconds after stimulus onset, while the post-stimulus undershoot is set
at around 15 seconds after onset. This function
was combined with each of the assumed designs using the convolution function. To account for possible
response delays and different slice acquisition times, we used in the three packages the first derivative of
the double gamma model, also known as the temporal derivative.
We did not incorporate physiological recordings to the analysis pipeline, as these were not available for
most of the datasets used.

We estimated the statistical maps in each package separately. AFNI, FSL and SPM use
Restricted Maximum Likelihood (ReML), where autocorrelation is estimated given the residuals from
an initial Ordinary Least Squares (OLS) model estimation. The ReML procedure then pre-whitens both the data and
the design matrix, and estimates the model.
We continued the analysis with the statistic maps corresponding to the t-test with null hypothesis being
that the full regression model without the canonical HRF explains as much variance as the
full regression model with the canonical HRF.
All three packages produced brain masks. The statistic maps in FSL and SPM
were produced within the brain mask only, while in AFNI the statistic maps were produced for
the entire volume. We masked the statistic maps from AFNI, FSL and SPM using the intersected brain
masks from FSL and SPM.
We did not confine the analyses to a gray matter mask, because autocorrelation is at strongest in gray
matter\cite{worsley2002general}. In other words, false positives caused by imperfect pre-whitening can
be expected to occur mainly in gray matter.
By default, AFNI and SPM produced t-statistic maps, while FSL produced both t-
and z-statistic maps. In order to transform
the t-statistic maps to z-statistic maps, we extracted the degrees of freedom from each analysis output.

Next, we performed multiple comparison correction in FSL for all the analyses, including for
those started in AFNI and SPM. First, we estimated the smoothness of the 
brain-masked 4-dimensional residual maps using the \texttt{smoothest} function in FSL. Knowing the
\texttt{DLH} parameter, which describes image roughness, and the number of voxels within the brain mask
(\texttt{VOLUME}), we then ran the \texttt{cluster} function in FSL on the z-statistic maps using a cluster
defining threshold of 3.09 and significance level of 5\%. This is the default multiple comparison
correction in FSL.
Finally, we applied previously saved MNI transformations to the binary maps which were showing the location
of the significant clusters.

\ \\
\noindent
\textbf{Results}

\begin{figure*}
   \begin{center}
	   \includegraphics[width=0.935\textwidth]{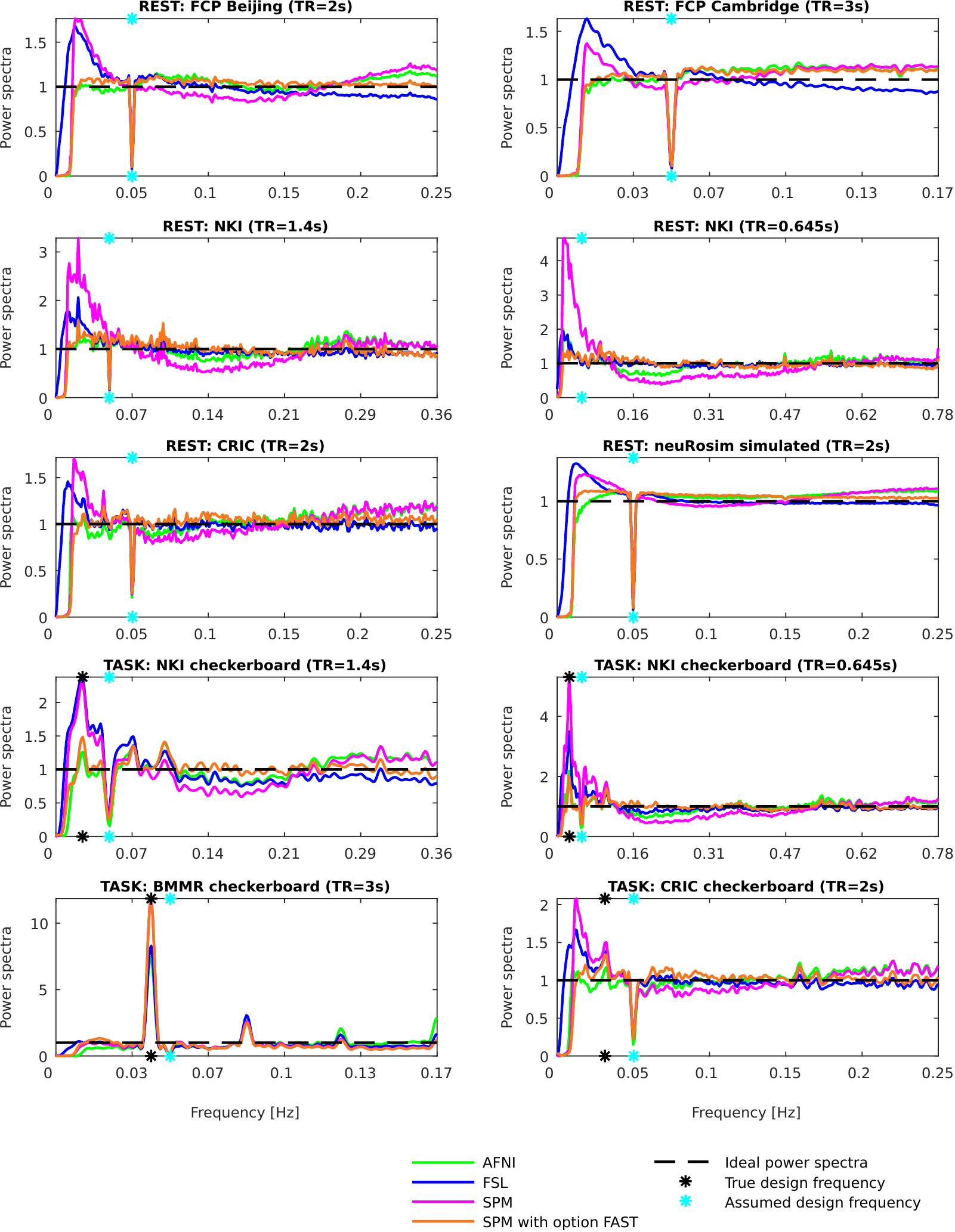}
   \end{center}
   \vspace{-2mm}
	\caption{Power spectra of the GLM residuals in native space averaged across brain voxels and across subjects
      for the assumed boxcar design
      of 10s of rest followed by 10s of stimulus presentation (``boxcar10'').
      The dips at 0.05 Hz are due to the assumed design period being 20s (10s + 10s).
      For some datasets, the dip is not seen as the assumed design frequency was not covered by one
      of the sampled frequencies.
      The frequencies on the x-axis go up to the Nyquist frequency, which is 0.5/TR.
      If after pre-whitening the residuals
      were white (as it is assumed), the power spectra would be flat. AFNI and SPM's alternative method:
      \texttt{FAST}, led to best whitening performance (most flat spectra). For FSL and SPM, there was
      substantial autocorrelated noise left after pre-whitening, particularly at low frequencies.}
   \label{power_spectra_8}
\end{figure*}

\noindent
\textbf{Whitening performance of AFNI, FSL and SPM.}
To investigate the whitening performance resulting from the use of noise models in AFNI, FSL and SPM, we plotted
the power spectra of the GLM residuals.
Figure~\ref{power_spectra_8} shows the power spectra averaged across
all brain voxels and subjects for smoothing of 8~mm and assumed boxcar design of 10s of rest
followed by 10s of stimulus presentation. The statistical inference in AFNI, FSL and SPM relies on the assumption
that the residuals after pre-whitening are white. For white residuals, the power spectra should be flat.
However, for all the datasets and all the packages, there was some visible structure. The strongest
artefacts were visible for FSL and SPM at low frequencies.
At high frequencies, power spectra
from \texttt{FAST} were closer to 1 than power spectra from the other methods.
Figure~\ref{power_spectra_8} does not show respiratory spikes which one could expect to
see.~This is because the figure refers to averages across subjects. We observed respiratory
spikes when analyzing power spectra for single subjects (not shown).

\begin{figure*}[t!]
   \includegraphics[height=18.8cm]{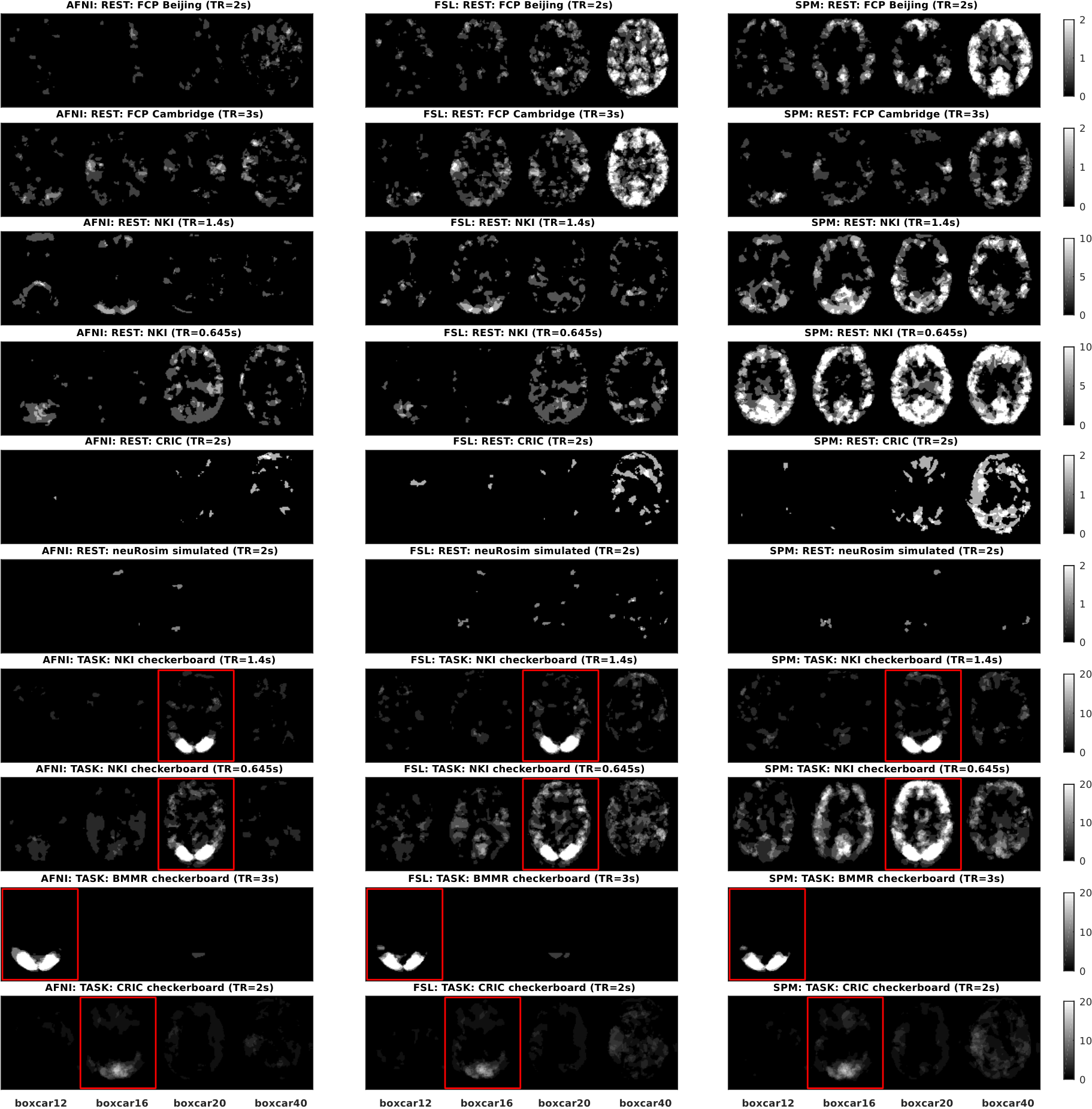}
	\caption{Spatial distribution of significant clusters in AFNI (left), FSL (middle) and SPM
      (right) for different assumed experimental designs. Scale refers to the percentage of subjects
      where significant activation was detected at the given voxel. The red boxes indicate the true designs
      (for task data).
      Resting state data was used as null data. Thus, low numbers of significant voxels were a desirable
      outcome, as it was suggesting high specificity.
      Task data with assumed wrong designs was used as null data too. Thus, large positive
      differences between the true design and the wrong designs were a desirable outcome.
      The clearest cut between the true and the wrong/dummy designs was obtained with AFNI's noise model.
      \texttt{FAST} performed similarly to AFNI's noise model (not shown).}
   \label{sp_dist}
\end{figure*}

\ \\
\noindent
\textbf{Resulting specificity-sensitivity trade-offs.}
In order to investigate the impact of the whitening performance on first level results, we analyzed the
spatial distribution of significant clusters in AFNI, FSL and SPM.
Figure~\ref{sp_dist} shows an exemplary axial slice in the MNI space for 8~mm smoothing. It was made
through the imposition of subjects' binarized significance masks on each
other. Scale refers to the percentage of subjects within a dataset where significant activation was
detected at the given voxel. The x-axis corresponds to four assumed designs.
Resting state data was used as null data. Thus, low numbers of significant
voxels were a desirable outcome, as this was suggesting high specificity. Task data with assumed wrong designs
was used as null data too. Thus, clear differences between the true design (indicated with red boxes)
and the wrong designs were a desirable outcome. For FSL and SPM, often the relationship between lower
assumed design frequency (``boxcar40'' vs. ``boxcar12'') and an increased number of significant voxels
was visible, in particular for the resting state datasets: ``FCP Beijing'', ``FCP Cambridge'' and ``CRIC''.
For null data, significant clusters in AFNI were scattered primarily within gray matter.
For FSL and SPM, many significant clusters were found in
the posterior cingulate cortex, while
most of the remaining significant clusters were scattered within gray matter across the
brain.~False positives in gray matter occur due to the stronger positive autocorrelation in this
tissue type compared to white matter\cite{worsley2002general}.
For the task datasets: ``NKI checkerboard TR=1.4s'', ``NKI checkerboard TR=0.645s'',
``BMMR checkerboard'' and ``CRIC checkerboard'' tested with the true designs, the majority of significant
clusters were located in the visual cortex.
This resulted from the use of visual experimental designs for the fMRI task.
For the impaired consciousness patients (``CRIC''), the registrations to MNI space were
imperfect, as the brains were often deformed.

\ \\
\noindent
\textbf{Additional comparison approaches.}
The above analysis referred to the spatial distribution of significant clusters on an exemplary axial slice.
As the results can be confounded by the comparison approach, we additionally investigated two other comparison
approaches: the percentage of significant voxels and the positive rate.
Supplementary
Fig.~1 shows the average percentage of significant voxels across subjects in 10 datasets for
smoothing of 8~mm and for 16 assumed boxcar experimental designs. As more designs were considered, the
relationship between lower assumed design frequency and an increased percentage of significant
voxels in FSL and SPM (discussed before for Fig.~\ref{sp_dist}) was even more apparent.
This relationship was particularly interesting for the ``CRIC checkerboard'' dataset. When tested with the true
design, the percentage of significant voxels for AFNI, FSL, SPM and \texttt{FAST} was similar: 1.2\%, 1.2\%,
1.5\% and 1.3\%, respectively. However, AFNI and \texttt{FAST} returned
much lower percentages of significant voxels for the assumed wrong designs. For the assumed wrong
design ``40'', FSL and SPM returned on average a higher percentage of significant voxels than for the true design:
1.4\% and 2.2\%, respectively. Results for AFNI and \texttt{FAST} for the same design showed only 0.3\% and
0.4\% of significantly active voxels.

Overall, at an 8~mm smoothing level, AFNI and \texttt{FAST} outperformed FSL and SPM showing a lower average
percentage of significant voxels in tests with the wrong designs: on average across 10 datasets and across the wrong
designs, the average percentage of significant voxels was 0.4\% for AFNI, 0.9\% for FSL, 1.9\% for SPM and
0.4\% for \texttt{FAST}.

As multiple comparison correction depends on the smoothness level of the residual maps, we also checked the
corresponding differences between AFNI, FSL and SPM. The residual maps seemed to be similarly smooth. At
an 8~mm smoothing level, the average geometric mean of the estimated FWHMs of the Gaussian distribution in x-, y-,
and z-dimensions across the
10 datasets and across the 16 assumed designs was 10.9~mm for AFNI, 10.3~mm for FSL, 12.0~mm for SPM and
11.8~mm for \texttt{FAST}.
Moreover, we investigated the percentage of voxels with z-statistic above 3.09. This value is the
99.9\% quantile of the standard normal distribution and is often used as the cluster defining threshold. For null
data, this percentage should be 0.1\%. The average percentage across the 10 datasets and across the
wrong designs was 0.6\% for AFNI, 1.2\% for FSL, 2.1\% for SPM and
0.4\% for \texttt{FAST}.

Supplementary Figs.~2-3 show the positive rate for smoothing of 4 and 8~mm. The
general patterns resemble those already discussed for the percentage of significant voxels, with AFNI and
\texttt{FAST} consistently returning lowest positive rates (familywise error rates) for resting state scans and task
scans tested with wrong designs. For task scans tested with the true designs, the positive rates for the different
pre-whitening methods were similar. The
black horizontal lines show the 5\% false positive rate, which is the expected proportion of scans with at least
one significant cluster if in reality there was no experimentally-induced signal in any of the subjects' brains. The
dashed horizontal lines are the confidence intervals for the proportion of false positives. These were calculated
knowing that variance of a Bernoulli($p$) distributed random variable is $p(1-p)$. Thus,
the confidence intervals were $0.05\pm\sqrt{0.05 \cdot 0.95 / n}$, with $n$ denoting the number of subjects in
the dataset.

Since smoothing implicitly affects the voxel size, we considered different smoothing kernel sizes.
We chose 4, 5 and 8~mm, as these are the defaults in AFNI, FSL and SPM.
No smoothing was also considered, as for 7T data this preprocessing step is sometimes
avoided\cite{walter2008high, polimeni2017analysis}.
With a wider smoothing kernel, the percentage of significant voxels increased (not shown), while the positive
rate decreased.
Differences between AFNI, FSL, SPM and \texttt{FAST} discussed above for the four comparison approaches and
smoothing of 8~mm were consistent across the four smoothing levels.

Further results are available from
\url{https://github.com/wiktorolszowy/fMRI_temporal_autocorrelation/tree/master/figures}

\ \\
\noindent
\textbf{Event-related design studies.}
In order to check if differences in autocorrelation modeling in AFNI, FSL and SPM lead to different first level
results for event-related design studies, we analyzed the CamCAN dataset. The task was a sensorimotor one with
visual and audio stimuli, to which the participants responded by pressing a button.
The design was based on an m-sequence\cite{buravcas2002efficient}.
Supplementary Fig.~4 shows (1) power spectra of the GLM residuals in native space averaged
across brain voxels and across subjects for the assumed true design (``E1''), (2) average percentage of
significant voxels for three wrong designs and the true design, (3) positive rate for the same
four designs, and (4) spatial
distribution of significant clusters for the assumed true design (``E1''). Only
smoothing of 8~mm was considered.
The dummy event-related design (``E2'') consisted of relative stimulus onset times generated from a uniform
distribution with limits 3s and 6s. The stimulus duration times were 0.1s.

For the assumed low-frequency design (``B2''), AFNI's autocorrelation modeling led to the lowest familywise error
rate as residuals from FSL and SPM again showed a lot of signal at low frequencies. However, residuals from SPM
tested with option \texttt{FAST} were similar at low frequencies to AFNI's residuals. As a result, the
familywise error rate was similar to AFNI. For high frequencies, power spectra from SPM tested with option
\texttt{FAST} were more closely
around 1 than power spectra corresponding to the standard three approaches (AFNI/FSL/SPM). For an event-related
design with very short stimulus duration times (around zero), residual positive autocorrelation at high frequencies
makes it difficult to distinguish the activation blocks from the rest blocks, as part of the experimentally-induced
signal is in the assumed rest blocks. This is what happened with AFNI and SPM. As their power spectra at high
frequencies were above 1, we observed for the true design a lower percentage of significant voxels compared to SPM tested with option
\texttt{FAST}. On the other hand, FSL's power spectra at high frequencies were below 1. As a result, FSL
decorrelated activation blocks from rest blocks possibly introducing negative autocorrelations at high
frequencies, leading to a higher percentage of significant voxels than SPM tested with option \texttt{FAST}.
Though we do not know the ground truth, we might expect that AFNI and SPM led for this event-related design
dataset to more false negatives than SPM with option \texttt{FAST}, while FSL led to more false positives.
Alternatively, FSL might have increased the statistic values above their nominal values for the truly but little
active voxels.

\ \\
\noindent
\textbf{Slice timing correction.}
As slice timing correction is an established preprocessing step, which often increases
sensitivity\cite{sladky2011slice},
we analyzed its impact on pre-whitening for two datasets for which we knew the acquisition order:
``CRIC checkerboard'' and ``CamCAN sensorimotor''.
``CRIC checkerboard'' scans were acquired with an
interleave acquisition starting with the second axial slice from the bottom (followed with fourth slice, etc.),
while ``CamCAN sensorimotor'' scans were acquired with a descending acquisition with the most upper axial slice
being scanned first. We considered only the true designs.
For the two datasets and for the four pre-whitening methods, slice timing correction changed the power spectra
of the GLM residuals in a very limited way (Supplementary Fig.~5). Regardless of whether slice timing correction
was performed or not, pre-whitening approaches from FSL and SPM left substantial positive autocorrelated noise at
low frequencies, while \texttt{FAST} led to even more flat power spectra than AFNI. We also investigated the average
percentage of significant voxels (Supplementary Table 1).
Slice timing correction changed the amount of significant activation only negligibly, with the exception of AFNI's
pre-whitening in the ``CamCAN sensorimotor'' scans.
In the latter case, the apparent sensitivity increase
(from 7.64\% to 13.45\% of the brain covered by significant clusters) was accompanied by power spectra of the GLM residuals
falling below 1 for the highest frequencies. This suggests negative autocorrelations were introduced at these frequencies,
which could have led to statistic values being on average above their nominal values.

\ \\
\noindent
\textbf{Group studies.}
To investigate the impact of pre-whitening on the group level, we performed via SPM random effects analyses
and via AFNI's \texttt{3dMEMA}\cite{chen2012fmri} we performed mixed effects analyses.
To be consistent with a previous study on group analyses\cite{eklund2016cluster}, we considered
one-sample t-test with sample size 20. For each dataset, we considered the first 20 subjects.
We exported coefficient maps and t-statistic maps (from which standard errors can be derived) following 8~mm
spatial smoothing and pre-whitening from AFNI, FSL, SPM and \texttt{FAST}.
Both for the random effects analyses and for the mixed effects analyses, we employed cluster inference with
cluster defining threshold of 0.001 and significance level of 5\%. Altogether, we performed 1312 group analyses:
2 (for random/mixed) $\times$ 4 (for pre-whitening) $\times$ ($10 \times 16+4)$
(for the first 10 datasets tested with 16 boxcar designs each and for the 11th dataset tested with four designs).
We found significant activation for 236 analyses, which we listed in Supplementary Table 2.

\begin{figure*}
   \includegraphics[width=\textwidth]{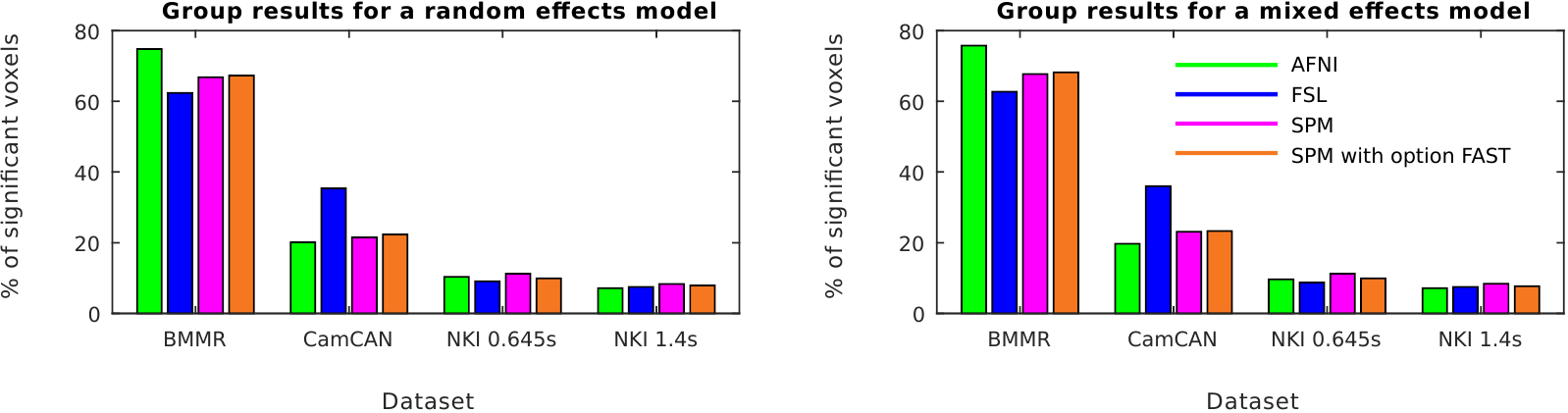}
   \caption{Group results for four task datasets with assumed true designs. Random effects analyses and
      mixed effects analyses led to only negligibly different average percentages of significant voxels.}
   \label{group_results}
\end{figure*}

For each combination of group analysis model and pre-whitening (2 $\times$ 4), we ran 164 analyses. As
five datasets were task datasets, 159 analyses ran on null data.
Supplementary Table 3 shows FWER for the random effects and mixed effects null data analyses, and for the four
pre-whitening approaches. On average, FWER for the mixed effects analyses was almost twice higher than FWER for
the random effects analyses. The use of AFNI's pre-whitening led to highest FWER,
while \texttt{FAST} led to lower FWER than the SPM's default approach.

Figure~\ref{group_results} shows the percentage of significant voxels for four task datasets
with assumed true designs. Results for the ``CRIC checkerboard'' dataset are not shown, as no significant
clusters were found at the group level. This occurred due to several of the subjects having deformed brains, which
led to the group brain mask not covering the primary visual cortex.
For the ``BMMR checkerboard'' dataset, the brain mask was limited mainly to the occipital lobe and the
percentage relates to the field of view that was used.
Both for the random effects analyses and for the mixed effects analyses, we observed little effect of
pre-whitening.
For task data tested with true designs, we found only negligible differences between the random effects analyses
and the mixed effects analyses.

Noteworthily, for the event-related task dataset ``CamCAN sensorimotor'' tested with the true design,
the use of \texttt{FAST} led to slightly higher amount of significant activation compared to the default SPM's method,
while FSL led to much higher amount of significant activation. This means that for this event-related design
dataset, the sensitivity differences from the first level analyses propagated to the second level.
This happened both for the random effects model and for the mixed effects model.

As the above results suggest that the use of standard error maps changes the group results in a very limited
way only, we investigated AFNI's \texttt{3dMEMA} by artificially re-scaling the t-statistic maps for one false
positive analysis: ``NKI rest (TR=1.4s)'' dataset with assumed design 36s off + 36s on. For each subject, we multiplied the
value of each voxel with 0.01, 0.1, 0.5, 2, 5 and 10.
We observed a surprising negative relationship between the magnitude of the t-statistic maps and the amount of
significant activation (Supplementary Table 4). Even when the t-statistics were extremely small (standard errors
100 times bigger compared to the original values), \texttt{3dMEMA} found significant
activation.
%In fact, for the four NKI datasets the average standard error for the null analyses was 0.22 for SPM and
%0.34 for \texttt{FAST}. Increased variance estimate should make it more difficult to detect significant activation.}

\ \\
\noindent
\textbf{Discussion}

\noindent
In the case of FSL and SPM for the datasets ``FCP Beijing'', ``FCP Cambridge'', ``CRIC RS'' and
``CRIC checkerboard'', there was a clear relationship between lower assumed
design frequency and an increased percentage of significant voxels.
This relationship exists when positive autocorrelation is not removed from the data\cite{purdon1998effect}.
%This phenomenon is caused by the spurious signal spillage. If during the assumed activation
%period the noise process spuriously takes high values and the assumed design frequency is high, due to
%the residual positive autocorrelation we can expect higher signal values during the beginning of the assumed
%rest period. Thus, it will be difficult to distinguish the assumed activation period from the assumed rest period,
%and the spuriously high signal during the former period will likely not result in detected significance.
%On the other hand, if such a spuriously high signal occurs in the middle of a long assumed activation period,
%there will be enough time for the signal to return to its baseline level, so that there will be a larger difference
%between the mean signal during the assumed activation period and the mean signal during the assumed rest period.
%As a result, detection of significant activation will be more likely.
Autocorrelated processes show increasing variances at lower frequencies. Thus, when
the frequency of the assumed design decreases, the mismatch between the true autocorrelated residual variance and
the incorrectly estimated white noise variance grows. In this mismatch, the variance is underestimated, which
results in a larger number of false positives.

An interesting case was the checkerboard experiment conducted with impaired consciousness
patients, where FSL and SPM found a higher percentage of significant voxels for the design with the assumed lowest
design frequency than for the true design.
As this subject population was unusual, one might suspect weaker or inconsistent
response to the stimulus. However, positive rates for this experiment for the true design
were all around 50\%, substantially above other assumed designs.

Compared to FSL and SPM, the use of AFNI's and \texttt{FAST} noise models for task datasets resulted in larger
differences between the true design and the wrong designs in the first level results. This occurred because of
more accurate autocorrelation modeling in AFNI and in \texttt{FAST}. In our analyses, FSL and SPM left a substantial
part of the autocorrelated noise in the data and the statistics were biased. For none of the pre-whitening
approaches were the positive rates around
5\%, which was the significance level used in the cluster inference. This is
likely due to imperfect cluster inference in FSL. High familywise error rates in first level FSL analyses
were already reported\cite{eklund2015empirically}. In our study the familywise error rate following the use of
AFNI's and \texttt{FAST} noise models was consistently lower than the familywise error rate following the use of
FSL's and SPM's noise models. Opposed to the average percentage of significant voxels, high familywise error rate
directly points to problems in the modeling of many subjects.

The highly significant responses for the NKI datasets are in line with previous findings\cite{eklund2012does}, where it was shown
that for fMRI scans with short TR it is more likely to detect significant activation. The NKI scans that we
considered had TR of 0.645s and 1.4s, in both cases much shorter than the usual repetition times. Such short
repetition times are now possible due to multiband sequences\cite{larkman2001use}.
The shorter the TR, the higher the correlations between adjacent time points\cite{purdon1998effect}.
If positive autocorrelation in the data is higher than the estimated level,
then false positive rates will increase. The former study\cite{eklund2012does} only referred to SPM.
In addition to the previous study, we observed that the familywise error rate for short TRs
was substantially lower in FSL than in SPM, though still much higher than for resting state scans at TR=2s
(``FCP Beijing'' and ``CRIC RS''). FSL models autocorrelation more flexibly than SPM, which seems to be
confirmed by our study. For short TRs, AFNI's performance deteriorated too, as autocorrelation spans much
more than one TR and an ARMA(1,1) noise model can only partially capture it.

Apart from the different TRs, we analyzed the impact of spatial smoothing.
If more smoothing is applied, the signal from gray matter will be often mixed with the signal from
white matter. As autocorrelation in white matter is lower than in gray matter\cite{worsley2002general},
autocorrelation in a primarily gray matter voxel will likely decrease following stronger smoothing.
The observed relationships of the percentage of significant voxels and of the positive rate from the smoothing
level can be surprising, as random field theory is believed to account for different levels of data smoothness.
The relationship for the positive rate (familywise error rate) was already
known\cite{eklund2012does, eklund2015empirically}. The impact of smoothing and spatial resolution was
investigated in a number of previous studies\cite{geissler2005influence, weibull2008investigation, mueller2017commentary}.
We considered smoothing only as a confounder. Importantly, for all four levels of smoothing,
AFNI and \texttt{FAST} outperformed FSL and SPM.

Our results confirm Lenoski et al.\cite{lenoski2008performance}, insofar as our study also showed problems
with SPM's default pre-whitening.
Interestingly, Eklund et al.\cite{eklund2015empirically} already compared AFNI, FSL and SPM
in the context of first level fMRI analyses. AFNI resulted in substantially lower false positive rates
than FSL and slightly lower false positive rates than SPM. We also observed lowest false positive rates for AFNI.
Opposed to that study\cite{eklund2015empirically}, which compared the packages in their entirety, we compared the packages
only with regard to pre-whitening. It is possible that pre-whitening is the most crucial single difference between
AFNI, FSL and SPM, and that the relationships described by Eklund et al.\cite{eklund2015empirically} would look completely
different if AFNI, FSL and SPM employed the same pre-whitening. For one dataset, Eklund et al.\cite{eklund2015empirically}
also observed that SPM led to worst whitening performance.

The differences in first level results between AFNI, FSL and SPM which we found could have been smaller if
physiological recordings had been modeled. The modeling of physiological noise is known to improve whitening
performance, particularly for short TRs\cite{lund2006non, BOLLMANN2018152, corbin2018accurate}. Unfortunately,
cardiac and
respiratory signals are not always acquired in fMRI studies. Even less often are the physiological recordings
incorporated to the analysis pipeline. Interestingly, a recent report suggested that the FSL's tool ICA FIX
applied to task data can successfully remove most of the physiological noise\cite{eklund2018cluster}. This was
shown to lower the familywise error rate.

In our main analysis pipeline we did not perform slice timing correction.
For two datasets, we additionally considered slice timing correction and observed very
similar first level results compared to the case without slice timing correction.
The observed little effect of slice timing correction is likely a result of the temporal derivative being
modeled within the GLM framework. This way a large part of the slice timing variation might have been captured
without specifying the exact slice timing.
For the only case where slice timing correction led to noticeably higher amount of significant activation, we
observed negative autocorrelations at high frequencies in the GLM residuals. If one did not see the power spectra
of the GLM residuals, slice timing correction in this case could be thought to directly increase
sensitivity, while in fact pre-whitening confounded the comparison.

FSL is the only package with a benchmarking paper of its pre-whitening approach\cite{woolrich2001temporal}.
The study employed data corresponding to two fMRI protocols. For one
protocol TR was 1.5s, while for the other protocol TR was 3s. For both protocols, the voxel size was 4x4x7 mm$^3$.
These were large voxels. We suspect that the FSL's pre-whitening approach could have been overfitted to this data.
Regarding SPM, pre-whitening with simple global noise models was found to result in profound
bias in at least two previous studies\cite{friston2000smooth, lenoski2008performance}.
SPM's default is a simple global noise model. However, SPM's problems could be partially related to
the estimation procedure. Firstly, the estimation is approximative as it uses a Taylor
expansion\cite{friston2002classical}. Secondly, the estimation is based on a subset of the voxels. Only voxels
with $p<0.001$ following inference with no pre-whitening are selected. This means that the estimation strongly
depends both on the TR and on the experimental design\cite{purdon1998effect}.

If the second level analysis is performed with a
random effects model, the standard error maps are not used. Thus, random effects models like the summary
statistic approach in SPM should not be affected by imperfect pre-whitening\cite{friston2005mixed}. On the
other hand, residual positive autocorrelated noise decreases the signal differences between the activation
blocks and the rest blocks. This is relevant for event-related designs.
Bias from confounded coefficient maps can be expected to propagate to the group level.
We showed that pre-whitening indeed confounds group analyses performed with a random
effects model.
However, more relevant is the case of mixed effects analyses, for example when using \texttt{3dMEMA} in
AFNI\cite{chen2012fmri} or \texttt{FLAME} in FSL\cite{woolrich2004multilevel}. These approaches additionally
employ standard error maps, which are directly confounded by imperfect pre-whitening. Bias in mixed effects fMRI
analyses resulting from non-white noise at the first level was already reported\cite{bianciardi2004evaluation}.
Surprisingly, we did not observe pre-whitening-induced specificity problems for analyses using \texttt{3dMEMA}, including for very
short TRs. While this means that imperfect pre-whitening does not meaningfully affect group results when using
\texttt{3dMEMA}, we wonder why the AFNI's mixed effects model makes so little use of the standard error maps.
For task datasets tested with true designs, the results from random effects analyses differed very little
compared to \texttt{3dMEMA} results.
Furthermore, we observed for \texttt{3dMEMA} a worrying negative relationship between the magnitude of the
t-statistic maps and the amount of significant activation. This is particularly surprising given that subject
heterogeneity in that analysis was kept constant.
\texttt{FLAME} was also shown to have similar sensitivity compared to random effects analyses\cite{mumford2009simple}.
However, mixed effects models should be more optimal than random effects models as they employ more information.
Although group analysis modeling in task fMRI studies needs to be investigated further, it is beyond the scope
of this paper. As mixed effects models employ standard error maps, bias in them should be avoided.

Problematically, for resting state data treated as task data, it is possible to observe activation both in the
posterior cingulate cortex and in the frontal cortex,
since these regions belong to the default mode network\cite{raichle2001default}.
In fact, in Supplementary Fig.~18 in Eklund et al.~2016\cite{eklund2016cluster} the spatial distribution plots of
significant clusters indicate
that the significant clusters appeared mainly in the posterior cingulate cortex, even though the assumed
design for that analysis was a randomized event-related design. The rest activity in
these regions can occur at different frequencies and can underlie different patterns\cite{stark2001zero}.
Thus, resting state data is not perfect null data for task fMRI analyses, especially if
one uses an approach where a subject with one small cluster in the posterior cingulate
cortex enters an analysis with the same weight as a subject with a number of large clusters spread throughout
the entire brain.
Task fMRI data is not perfect null data either, as an assumed wrong
design might be confounded by the
underlying true design. For simulated data, a consensus is needed how to model autocorrelation, spatial
dependencies, physiological noise, scanner-dependent low-frequency drifts and head motion. Some of the current
simulation toolboxes\cite{welvaert2014review} enable the modeling of all these aspects of fMRI data,
but as the later analyses might heavily depend on the specific choice of parameters, more work is needed to
understand how the different sources of noise influence each other.
In our study, results for simulated resting state data were substantially different compared to acquired real
resting state scans. In particular, the percentage of significant voxels for the simulated
data was much
lower, indicating that the simulated data did not appropriately correspond to the underlying brain physiology.
Considering resting state data where the posterior cingulate cortex and the frontal cortex are masked out could
be an alternative null. Because there is no perfect fMRI null data, we used
both resting state data with assumed dummy designs and task data with assumed wrong designs.
Results for both approaches coincided.

Unfortunately, although the vast majority of task fMRI analyses is conducted with linear regression, the
popular analysis packages do not provide diagnostic plots. For old versions of SPM, the external toolbox
\texttt{SPMd} generated them\cite{luo2003diagnosis}. It provided a lot of information, which paradoxically
could have limited its popularity.
We believe that task fMRI analyses would strongly benefit if AFNI, FSL
and SPM provided some basic diagnostic plots. This way the investigator would be aware, for example,
of residual autocorrelated noise in the GLM residuals.
We provide a simple MATLAB tool (GitHub: \path{plot_power_spectra_of_GLM_residuals.m}) for the fMRI researchers
to check if their analyses might be affected by imperfect pre-whitening.

To conclude, we showed that AFNI and SPM tested with option
\texttt{FAST} had the best whitening performance, followed by FSL and SPM. Pre-whitening in FSL and SPM left
substantial residual autocorrelated noise in the data, primarily at low frequencies.
Though the problems were most severe for short repetition times, different fMRI protocols were affected.
We showed that the residual autocorrelated noise led to heavily confounded first level
results. Low-frequency boxcar designs were affected the most. Due to better whitening performance,
it was much easier to distinguish the assumed true experimental design from the assumed wrong experimental
designs with AFNI and \texttt{FAST} than with FSL and SPM. This suggests superior specificity-sensitivity trade-off
resulting from the use of AFNI's and \texttt{FAST} noise models.
False negatives can occur when the design is event related and there is residual positive autocorrelated noise
at high frequencies. In our analyses, such false negatives propagated to the group level both when using a
random effects model and a mixed effects model, although only to a small extent.
Surprisingly, pre-whitening-induced false positives did
not propagate to the group level when using the mixed
effects model \texttt{3dMEMA}. Our results suggest that \texttt{3dMEMA} makes very little use of the standard error
maps and does not differ much from the SPM's random effects model.

Results derived from FSL could be made more robust if a different autocorrelation model
was applied. However, currently there is no alternative pre-whitening approach in FSL. For SPM, our findings
support more widespread use of the \texttt{FAST} method.

\ \\
{\setstretch{0.75}
{\scriptsize
\noindent
\textbf{Data availability.} FCP\cite{biswal2010toward}, NKI\cite{nooner2012nki} and
CamCAN data\cite{shafto2014cambridge} are publicly shared anonymized data. CRIC and BMMR scans can
be obtained from us upon request. The simulated data can be generated again using our GitHub
script \texttt{simulate\_4D.R}.} \par }

\ \\
\noindent
\textbf{References}

%\bibliographystyle{ieeetr}
%-to remove the heading
%-https://tex.stackexchange.com/questions/22645/hiding-the-title-of-the-bibliography
\renewcommand{\section}[2]{}
%-to remove space between entries
%-https://tex.stackexchange.com/questions/93859/condense-the-space-between-bibliographic-entries
\setlength{\bibsep}{0pt plus 0.3ex}
{\scriptsize
%\bibliography{Olszowy_fMRI_autocorrelation}
 }

\ \\
\textbf{Acknowledgements}

{\setstretch{0.75}
{\scriptsize
\noindent
We would like to thank Micha\l{} Kosicki, Paul Browne, Anders Eklund, Thomas Nichols, Karl Friston,
Richard Reynolds, Carsten Allefeld, Paola Finoia,
Adrian Carpenter, Alison Sleigh, Gang Chen, and Guillaume Flandin for much valuable advice.
Furthermore, we would like to thank the James S. McDonnell
Foundation for funding the image acquisitions of the Cambridge Research into Impaired Consciousness (CRIC) group,
and the CRIC group for sharing their data.
Oliver Speck, Michael Hoffmann and Aini Ismafairus Abd Hamid from the Otto von Guericke University provided us with the
7T data. We also thank the Neuroimaging Informatics Tools and Resources Clearinghouse and all of the researchers
who have contributed with data to the 1,000 Functional Connectomes Project and to the
Enhanced Nathan Kline Institute - Rockland Sample.
W.O. was in receipt of scholarships from the Cambridge Trust and from the Mateusz B. Grabowski Fund.
Also, W.O. was supported by the Guarantors of Brain.} \par }
%-'par' a trick for setstretch to work
%-https://tex.stackexchange.com/questions/176983/setstretch-command-does-not-work

\ \\
\textbf{Author contributions}

{\setstretch{0.75}
{\scriptsize
\noindent
W.O., J.A., and G.B.W. designed the study; W.O. conducted the study;
W.O., J.A., C.R., and G.B.W. analyzed the data; W.O. wrote the paper.} \par }

\ \\
\noindent
\textbf{Additional information}

{\setstretch{0.75}
\noindent
{\scriptsize \textbf{Competing interests:} The authors declare no competing interests.} \par }

\end{document}